# Simple-layered high mobility field effect heterostructured two-dimensional electron device

R.L. Willett, L.N. Pfeiffer, K.W. West
Bell Laboratories, Lucent Technologies

ABSTRACT:
We present a two-dimensional electron heterostructure field effect device of simplistic design and ease of fabrication that displays high mobility electron transport. This is accomplished using a high efficacy contacting scheme and simple metallic overlapping gate, obviating dopant layers. The resultant devices demonstrate adjustable electron densities and mobilities larger than $8 \times 10^6$ cm$^2$/V-sec at the highest densities of $2.4 \times 10^{11}$/cm$^2$. This device type provides an experimental avenue for studying electron correlations and may answer demands for routine fabrication of practical HEMTs.

High mobility two-dimensional electron systems (2DES) are of practical significance [1] and have had substantial impact on fundamental studies of electron-electron interactions [2]. Record high mobility systems [3] have employed modulation doped AlGaAs/GaAs heterostructures [4], utilizing remoteness of the ionized impurities to the 2D electron conduction layer. Pursuing a different route to 2D channel population, experimental devices have been produced that do away with the modulated Si dopant [5], instead field effecting the 2D electron channel using a heavily doped top-gate MBE deposited layer. These structures have limitations in their fabrication due to necessary self-alignment of the surface gate to the edges of the source-drain contacting materials.

In this letter we present a field effect device of simplistic design that overcomes this alignment issue, adding ease of fabrication while maintaining high mobility. The device makes use of a high efficacy contacting scheme and simple metallic overlapping gate. The devices demonstrate adjustable electron densities from 0.5 to 2.4 x $10^{11}$/cm$^2$ and mobilities up to $8.3 \times 10^6$ cm$^2$/V-sec. This device type provides an experimental avenue for studying electron correlations and may answer demands for routine fabrication of practical high electron mobility transistors (HEMTs).

In reviewing sample construction the critical principal for operation of these devices is to accomplish vertical diffusion of the contact *perimeter* to the interface layer where the electrons will reside. Only with this exposure of the diffused contact can the field effect process work.

Components of the field-effect device are shown schematically in Figure 1. A simple single interface heterostructure is produced through molecular beam epitaxy of Al$_x$Ga$_{1-x}$As and GaAs. The interface supporting the field-effected two-dimensional (2D) electron layer is the interface between these materials, and the Al$_x$Ga$_{1-x}$As layer thickness has been varied, using either 200nm or 100nm. Devices both with and without a thin GaAs cap layer of 50Å thickness have been produced with comparable ultimate performance. The aluminum concentration x is controlled and its fraction is important to the device operation. A standard mesa is etched in the sample surface with wet etching to a depth of 200nm using a solution with volume ratios of 100/10/2 for $H_2O/NH_4OH/H_2O_2$ where the stock $H_2O_2$ concentration is 30%. The mesa dimension is roughly 500µm on a side and facilitates van der Pauw transport measurements [6].

Proper contact diffusion into the perimeter of the mesa is critical. A layered contact of Ni/Au/Ge/Ni (4nm/200nm/100nm/800nm), starting with the 4nm Ni layer, is deposited on photolithographically defined patterns as shown in Figure 1. Note the overlap of the contact with the mesa for each contact is roughly 10 µm x 30µm. Contacts are diffused into the mesa using a rapid thermal annealer over a sustained temperature of 450C for 10 minutes.

An insulating layer of $Si_3N_4$ is then deposited over the entire sample surface following precaution to protect the outer contact pads by placing removable indium lumps on the pads. Standard plasma enhanced chemical vapor deposition (PECVD) is performed using a gas source reaction chamber. The $Si_3N_4$ layer thickness used in these devices was 120nm. A metallic top gate is then overlayed on the entire 2D mesa and the contacts. See Figure 1. This photolithographically defined gate is 30nm of aluminum, thermally evaporated at a rate of ~0.5nm/sec for 10 seconds, then >1.5nm/sec to achieve the target thickness. All results displayed here are from samples A-D with $Al_xGa_{1-x}As$ layer thickness of 100nm, prepared as described above, and with either 10% or 24 % x.

A positive bias voltage applied between the top gate and the contacts serves to draw carriers into the $Al_xGa_{1-x}As$ /GaAs interface. Once populated the contacts can be used in standard transport provided that one of the contacts remains a reference pole for the applied bias. By varying the applied voltage the density of carriers can be adjusted within sample parameter constraints.

d.c. magneto-transport for both a low and a high aluminum concentration x sample are displayed in Figure 2. Well-developed fractional quantum Hall states are apparent at 290mK. These simple devices are clearly able to support correlated electron states at modest temperatures.

Further experiments establish mobility and density for different sample aluminum concentrations and applied gate voltages. Figure 3 shows the resultant densities for a range of applied gate voltages and for 10% and 24% aluminum concentrations. Both have initial population of the 2D channel for gate voltages near 0.7V and both show density increase at approximately $0.5 \times 10^{11} cm^{-2}/V$. The density shows a maximum value for each aluminum concentraion such that higher applied gate voltage does not produce higher density. The maximum density achieved for 10% Al is near $0.6 \times 10^{11} cm^{-2}$ while that for 24% is near $2.4 \times 10^{11} cm^{-2}$. The maximum density does not represent the consequence of breakdown of the gate voltage to the 2D channel as current monitoring shows no significant leakage.

Mobility values for the different applied voltages are displayed in Figure 4. Both show mobility increase with density increase as expected for an expanding Fermi surface radius. The 10% aluminum concentration samples displayed some increase in measured mobility at gate voltages greater than 1.1V even though the density saturated for higher gate bias: this may indicate a parallel channel within the sample.

The devices described here are able to achieve substantial mobilities over a range of densities with the mechanisms of 2D electron channel population based on the layer properties. The 2D layers turn on to nominal densities of roughly $0.2 \times 10^{11}/cm^2$ for an applied gate voltage of 0.7V regardless of the aluminum concentration x or the $Al_xGa_{1-x}As$ layer thickness: this voltage corresponds roughly to the voltage difference of the GaAs Fermi level for an undoped substrate and the bottom of the conduction band, in agreement with a mid-band-gap Fermi level. A maximum density is reached at each aluminum concentration for sufficiently high gate bias without evidence for gate leakage, and this is consistent with charge population at the $Al_xGa_{1-x}As$ /insulator interface: charge may be trapped at the interface with its population above the roll-off gate voltage value. Here the aluminum concentration x may determine the barrier height to charge loss from the 2D channel to the $Al_xGa_{1-x}As$/insulator interface, also resulting in a higher potential 2D channel density. The upper limit of aluminum concentration to produce high-density 2D channels is presently under study. While these measured device parameters have some correspondence to certain band properties of the materials, overall, a precise band picture of this structure is still open to determination.

To summarize the results, high mobility 2D electron gases able to demonstrate numerous states of the fractional quantum Hall effect are produced using a simple field effect configuration. Electron densities ranging from $0.2 \times 10^{11}/cm^2$ to $2.4 \times 10^{11}/cm^2$ are achievable and dependent upon the sample parameters, with mobilities in excess of $8 \times 10^6 cm^2/V\text{-sec}$ observed at the highest densities. These simple devices offer a scheme for producing tunable high mobility electron channels for fundamental correlated electron studies. Further work is examining fundamental properties of these devices and possible p-type performance.

FIGURE CAPTIONS:
Figure 1: Schematic drawing of heterostructure field effect device demonstrating the simple heterostructure layering and penetration of the contacts vertically to the AlGaAs/GaAs layer. Photomicrograph of a typical device with gate overlaying contacts and 500μm x 500μm mesa.

Figure 2: d.c. transport from two simple field effect devices samples A and B with Al concentrations of 10% and 24% respectively and with varied gate voltages resulting in different densities. Measurement temperature is 290mK.

Figure 3: Densities consequent to applied gate voltage for both the 10% and 24% aluminum concentration field effect devices C and D respectively at 290mK. The Al concentration determines the maximum achievable electron density.

Figure 4: Mobility consequent to applied gate voltage for 10% and 24% aluminum concentrations in devices C and D respectively, at 290mK. The inset displays mobility as a function of density for the 24% device D, with approximate mobility $\mu \sim n^{0.7}$ [3]; the straight line is a guide for the eye.

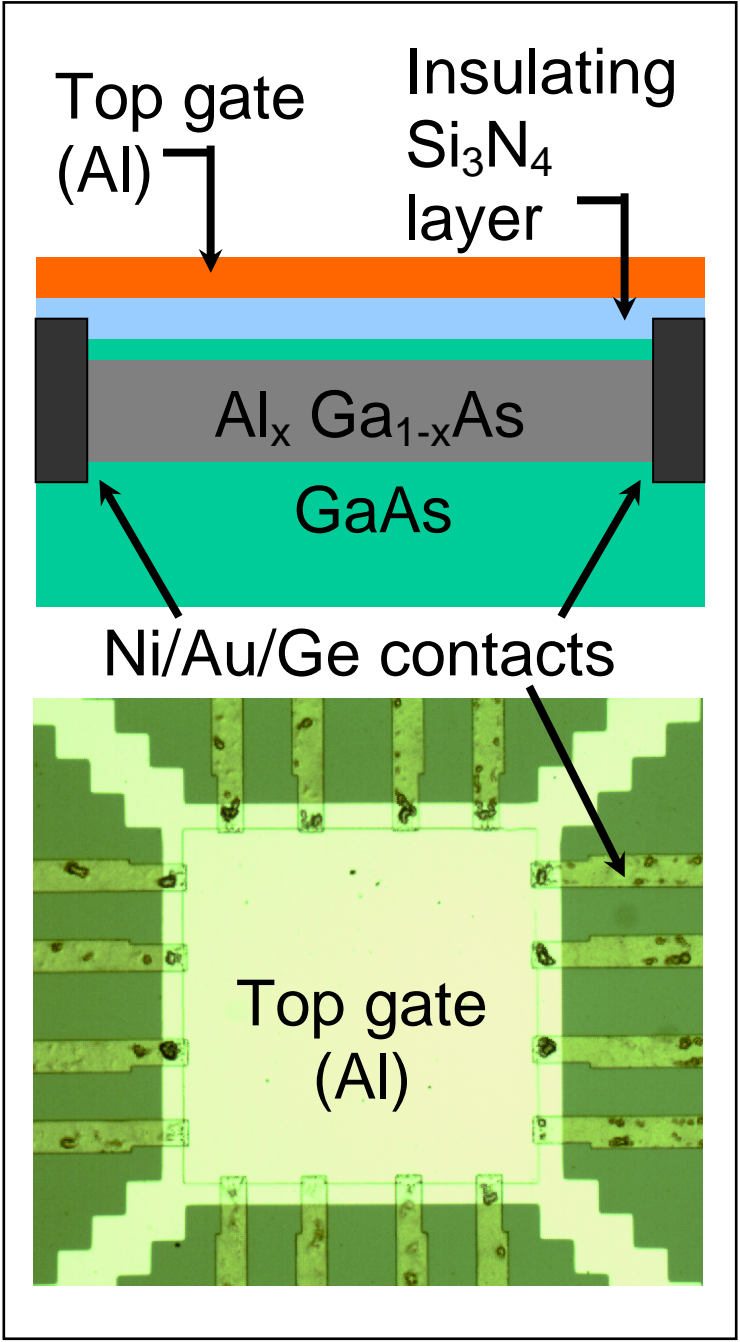

Figure 1

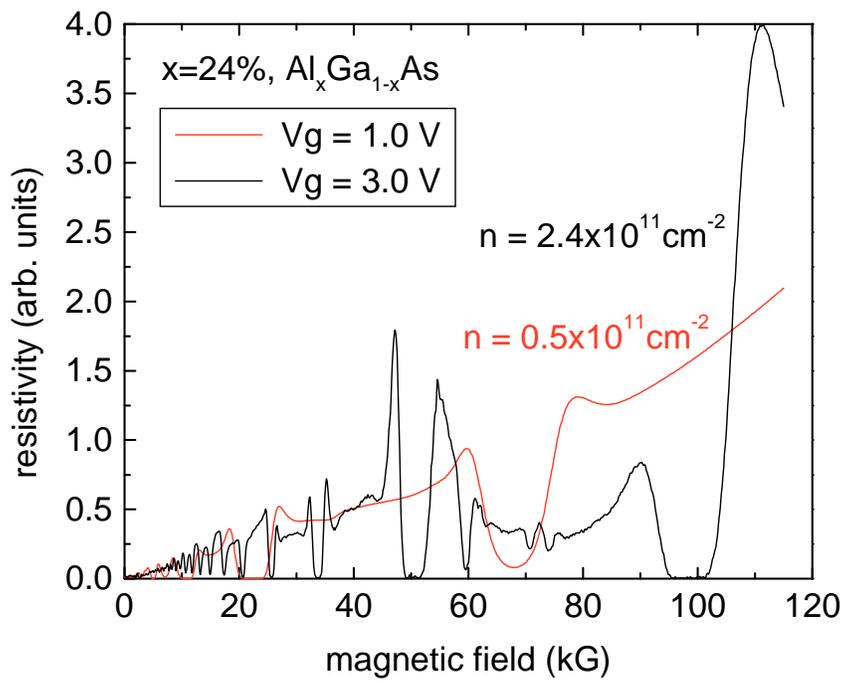
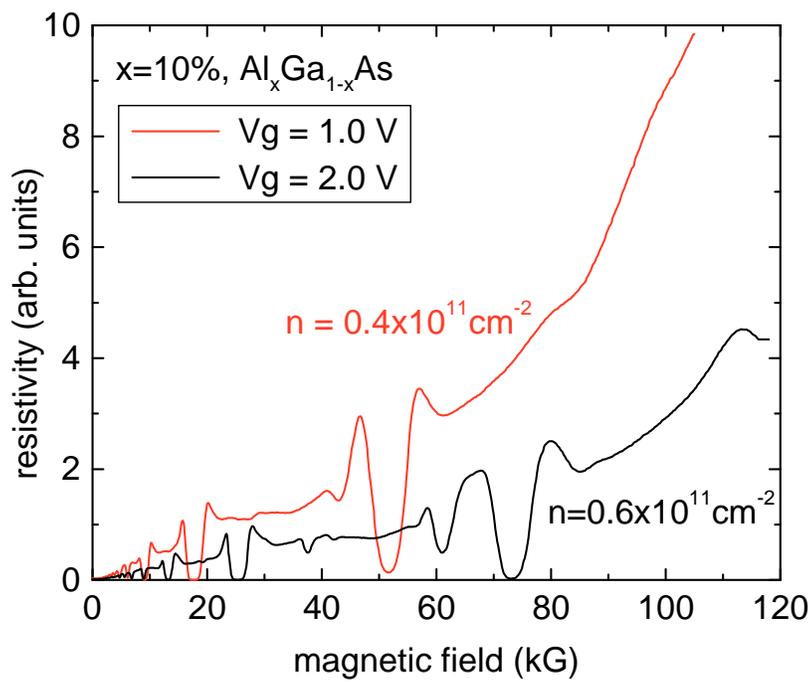

Figure 2

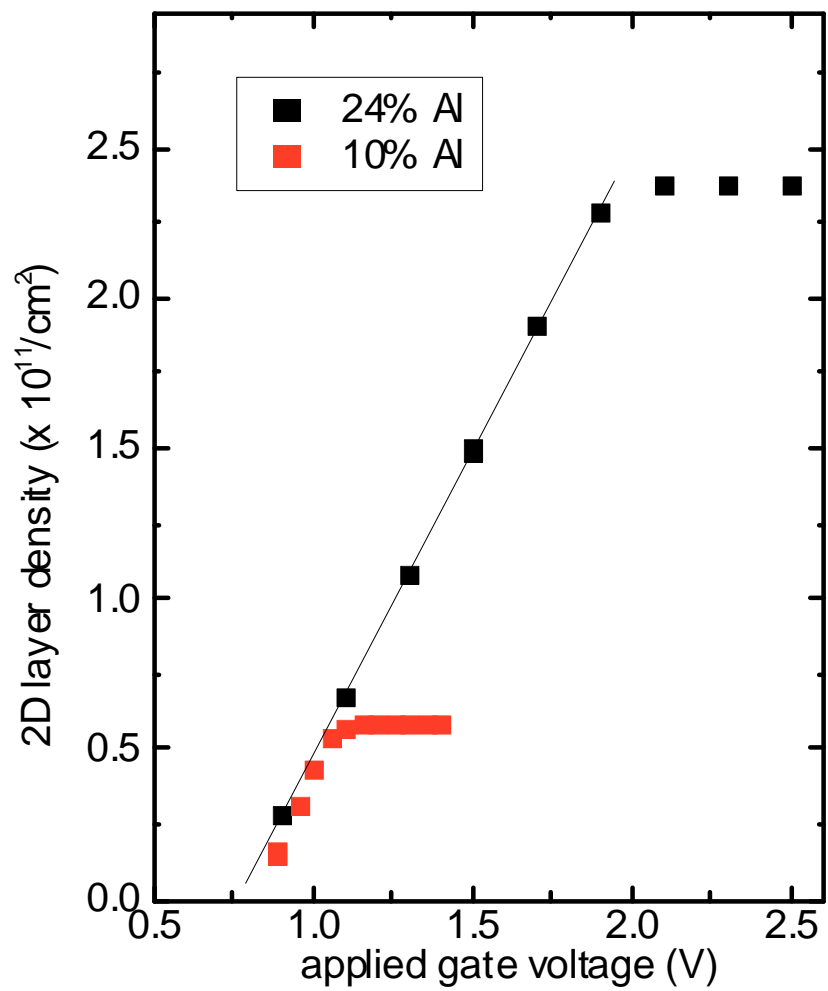

Figure 3

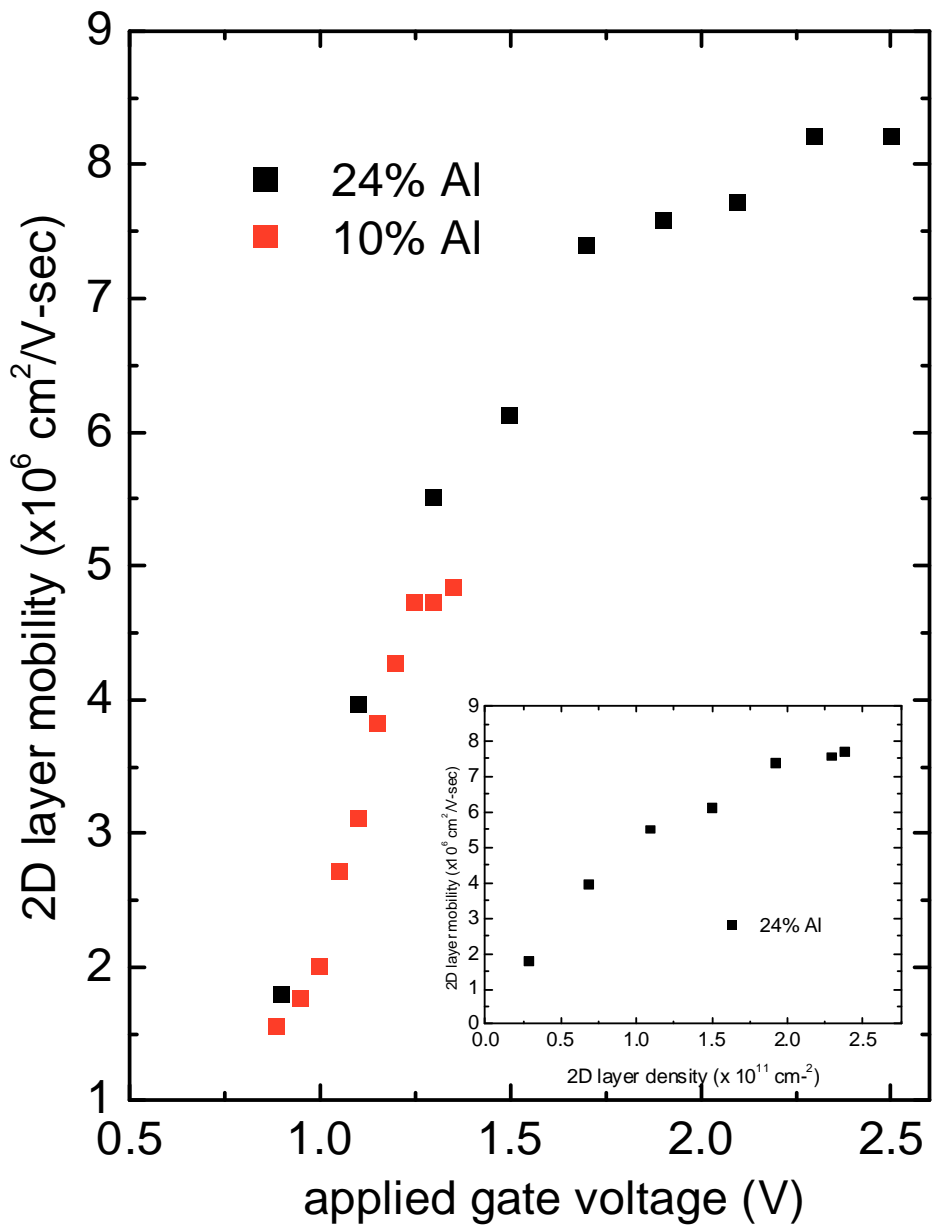

Figure 4